\documentclass[smallextended]{svjour3}      
\smartqed  
\usepackage{graphicx}
\usepackage[misc]{ifsym}
\usepackage{subfigure}
\usepackage{mathtools}
\usepackage{mathptmx}
\usepackage{mathrsfs}
\usepackage{amssymb}
\usepackage{amsbsy}
\usepackage{epsfig}
\usepackage{float}
\usepackage{amsmath}
\usepackage{color}
\usepackage{authblk}
\usepackage{stackrel}
\usepackage{cite}
\usepackage{url}
\usepackage{textcomp}
\usepackage{array}
\usepackage{float}

\begin{document}

\title{Sharing quantum steering among multiple Alices and Bobs via a two-qubit Werner state}
\titlerunning{Sharing quantum steering among multiple Alices and Bobs via a two-qubit Werner state}

\author{Xinhong Han\textsuperscript{1} \and Ya Xiao\textsuperscript{1,~\Letter}\and Huichao Qu\textsuperscript{1}\and Runhong He\textsuperscript{1}\and Xuan Fan\textsuperscript{1}\and Tian Qian \textsuperscript{1}\and Yongjian Gu \textsuperscript{1,~\Letter}
}
\institute{\Letter \quad Ya Xiao \\
	\setlength{\parindent}{2em} \indent xiaoya@ouc.edu.cn\\ \\
	\Letter \quad Yongjian Gu \\
	\setlength{\parindent}{2em} \indent yjgu@ouc.edu.cn\\ \\
	\textsuperscript{1} \quad Department of Physics, Ocean University of China, Qingdao 266100, People's Republic of China \\}
\authorrunning{X. Han et al.}
\date{Received: date / Accepted: date}
\maketitle

\begin{abstract}

Quantum steering, a type of quantum correlation with unique asymmetry, has important applications in asymmetric quantum information tasks. We consider a new quantum steering scenario in which one half of a two-qubit Werner state is sequentially measured by multiple Alices and the other half by multiple Bobs. We find that the maximum number of Alices who can share steering with a single Bob increases from 2 to 5 when the number of measurement settings $N$ increases from 2 to 16. Furthermore, we find a counterintuitive phenomenon that for a fixed $N$, at most 2 Alices can share steering with 2 Bobs, while 4 or more Alices are allowed to share steering with a single Bob. We further analyze the robustness of the steering sharing by calculating the required purity of the initial Werner state, the lower bound of which varies from 0.503(1) to 0.979(5). Finally, we show that our both-sides sequential steering sharing scheme can be applied to control the steering ability, even the steering direction, if an initial asymmetric state or asymmetric measurement is adopted. Our work gives insights into the diversity of steering sharing and can be extended to study the problems such as genuine multipartite quantum steering when the sequential unsharp measurement is applied. 
\end{abstract}

\keywords{Quantum steering \and Both sides unsharp measurement \and Werner state \and $N$-setting linear steering criterion}

\section{Introduction}
\label{intro}
Quantum steering was first proposed by Schr\"odinger in 1936 \cite{Schrodinge1936} in response to the EPR paradox \cite{einstein1935can}. However, it did not attract much attentions until 2007, when Wiseman \textit{et al.} reinterpreted quantum steering strictly from the operational view and even proposed some experimental criteria \cite{wiseman2007steering}. Ever since, the research of quantum steering has made great progress both in theory \cite{piani2015necessary,sun2017exploration} and experiment \cite{kocsis2015experimental,cavalcanti2016quantum,deng2017demonstration,zhao2020experimental,wollmann2020experimental}. In Wiseman's definition, quantum steering that logically intermediates between quantum entanglement and Bell nonlocality, describes the ability of one party, Alice, to nonlocally control the state of another party, Bob, even when Bob does not trust Alice's measurement apparatus, exhibiting unique asymmetric behavior  \cite{gallego2015resource,he2013genuine,xiao2017demonstration,uola2020quantum}. As an essential type of quantum correlations, quantum steering has great applications in quantum key distribution \cite{gehring2015implementation,walk2016experimental}, subchannel discrimination \cite{sun2018demonstration}, asymmetric quantum network \cite{cavalcanti2015detection}, randomness generation \cite{skrzypczyk2018maximal,guo2019experimental} and randomness certification \cite{curchod2017unbounded}.   
In the standard EPR steering tasks, $N$ entangled particles are separately distributed to $N$ different observers and each observer performs some projective (sharp) measurements to demonstrate her or his steerability. Since each observer is spatially separated, the non-signaling condition is strictly satisfied between different observers, i.e., the marginal probability distribution of each observer does not depend on the measurements of any other observers \cite{masanes2006general}. Due to the monogamy constraints, the number of observers who share quantum correlation via sharp measurement is limited \cite{coffman2000distributed, toner2006monogamy,reid2013monogamy,mal2017necessary}. Recently, a surprising result was reported by Silva \textit{et al.} that the number of observers sharing non-locality can be increased if the sequential weak (unsharp) measurement was employed, where the non-signaling condition is dropped \cite{silva2015multiple}. Their result later is confirmed by theoretical \cite{mal2016sharing,das2019facets,brown2020arbitrarily} as well as experimental works \cite{schiavon2017three,hu2018observation} and the sequential unsharp measurement strategy has been extended to study other types of quantum correlation \cite{bera2018witnessing,datta2018sharing,saha2019sharing}. It has shown that the maximum number of Alices who can simultaneously share steering with a single Bob can also beat the steering monogamy limits \cite{sasmal2018steering,shenoy2019unbounded,choi2020demonstration}.

However, all the steering sharing scenarios \cite{sasmal2018steering,choi2020demonstration} investigated till now have the following commonalities: the initial shared state is restricted to be the maximum entangled state, the sequential unsharp measurement is only adopted by one side, and the number of measurement settings is not more than 3. Thus some interesting questions raise: whether or not the steering correlation can be kept when the shared state is not pure any more? If there exist multiple observers on both sides, can multiple Alices steer multiple Bobs simultaneously? Compared to the single Bob case, do multiple Bobs make a difference?  And how many observers can share steering simultaneously if the number of measurement settings increases? 

In this work, we consider a more general sequential steering scenario featuring that unsharp measurements are sequentially performed on both sides. We investigate how many pairs of Alice and Bob can sequentially demonstrate steering in the above scenario when each party performs $N$-setting equal sharpness measurements. With the $N$-setting linear steering criterion  \cite{cavalcanti2009experimental}, we find no more than 5 Alices can steer a single Bob for a Werner state when $N$ increases from 2 to 16. Then we show how such sequential steering sharing scenarios tolerate the environmental noise and experimental imperfections by analyzing the useful sharpness measurement range of each observer and the minimum purity bound of the initial state. Furthermore, we explore the case when multiple Bobs involved in, reporting a counter-intuitive result that at most 2 Alices can steer 2 Bobs even the number of steering sharing observers larger than 4 in the single Bob case. Finally, we show that our scenario can be used to simulate quantum decoherence channels to effectively change the ability and direction of quantum steering. Our results not only reveal the rich structure of steering sharing but also can be applied to more general scenarios involving high dimension or genuine multipartite quantum steering  \cite{he2013genuine}.

\section{The both-sides sequential steering sharing scenario}
\begin{figure}[htbp]
	\begin{center}
		\includegraphics[width=10cm]{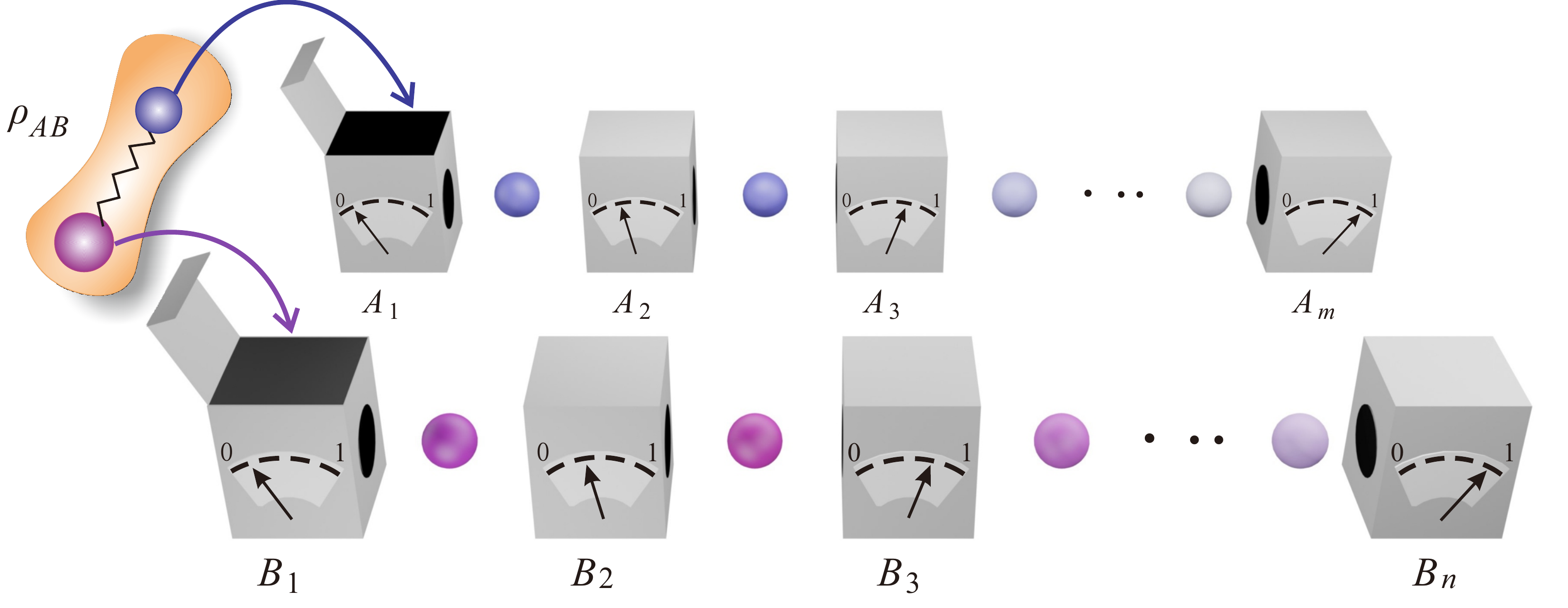}
		\caption{The scenario of steering sharing with multiple observers on both sides. A two-qubit entangled state is initially shared between a sequence of Alices and Bobs. Multiple Alices implement unsharp measurements on one part of the state successively, and multiple Bobs do similar operations on the other part.}
		\label{Fig 1}
	\end{center}
\end{figure}
A schematic of steering sharing scenario with both sides sequential unsharp measurements is shown in Fig.~\ref{Fig 1}. A pair of two-qubit entangled state  $\rho _{AB} $  is sent to multiple pairs of spatially separated observers. One of the qubits is accessed by $m$ Alices, say, $A_{1}$, $A_{2}$,..., $A_{m}$, while the other qubit is possessed by $n$ Bobs, say, $B_{1} $, $B_{2}  $,..., $B_{n}$.
To demonstrate the steering between multiple Alices and Bobs at the same time, all observers except the last Alice and Bob should perform unsharp measurements, otherwise, the steerability will be completely destroyed. For convenience, we assume that the sharpness of the $N$-setting measurements that each observer used is equal, which is denoted as $\lambda_{i}$ and $\eta_{p}$ for the $ i $-th Alice and the $ p $-th Bob respectively. Thus their corresponding $N$-setting measurements can be represented by  $\lbrace \hat{\Pi}^{\lambda_{i}}_{\vec{m}_{1}}, \hat{\Pi}^{\lambda_{i}}_{\vec{m}_{2}}$,..., $ \hat{\Pi}^{\lambda_{i}}_{\vec{m}_{N}}\rbrace $ and $\lbrace \hat{\Lambda}^{\eta_{p}}_{\vec{n}_{1}}, \hat{\Lambda}^{\eta_{p}}_{\vec{n}_{2}}$,..., $ \hat{\Lambda}^{\eta_{p}}_{\vec{n}_{N}}\rbrace $, where $\vec{m}_{k}$ and $\vec{n}_{k}$ represent the measurement directions with $k\in\lbrace 1 ,...,\, N \rbrace $, $i\in\lbrace 1 ,...,\, m \rbrace $, $p\in\lbrace 1 ,..., \, n \rbrace $, $\lambda_{i} \in [0,1]$ and $\eta_{p} \in [0,1]$. $\lambda_{i} (\eta_{p})\!=\!0$ corresponds to no measurement, $\lambda_{i} (\eta_{p})\!=\!1$ implies the measurement is sharp, and $\lambda_{i} (\eta_{p})\in(0,1)$ means it is unsharp. It has been demonstrated that an unsharp measurement is optimal when quality factor $ F $ and the precision $ G $ of the measurement satisfy the trade-off relation $ F^{2}+G^{2}=1 $ \cite{silva2015multiple}.
Here, each observer adopts the optimal measurement strategy.

In the first step, suppose $A_{1}$ wants to convince $B_{1}$ that she can remotely affect his state through local measurements. However, $B_{1}$ does not trust her, so he asks $A_{1}$ to perform a measurement along $\vec{m}_{k}$. After each run of experiment, $A_{1}$ sends $B_{1}$ the corresponding outcome $ a_{k}\in \lbrace 0,1 \rbrace$ and sends $A_{2}$ the post-measurement state, which can be described by the L\"uders ruler \cite{busch1986unsharp}
\begin{center}
	\begin{equation}
	\rho_{AB}\rightarrow(\!K^{\lambda_{1}}_{a_{k}\mid\vec{m}_{k}} \otimes I_{B} )\rho_{AB}(K^{\lambda_{1}\dagger}_{a_{k}\mid\vec{m}_{k}} \otimes I_{B}),
	\end{equation}
\end{center}
where $K^{\lambda_{1}}_{a_{k}\mid\vec{m}_{k}} K^{\lambda_{1}\dagger}_{a_{k}\mid\vec{m}_{k}}\!=\!\hat{\Pi}^{\lambda_{1}}_{a_{k}\mid\vec{m}_{k}}\!=\!(I_{A}+(-1)^{a_{k}} \lambda_{1}\,\vec{m}_{k}\cdot\vec{\sigma})/2$, $\vec{\sigma}\!=\!\{\sigma_x, \sigma_y, \sigma_z\}$ is the Pauli matrix, and $I_{A}$  $(I_{B})$ is the identity matrix. Repeating the process many times, when $A_{1}$ finishes all the $N$-setting measurements, $B_{1}$ will obtain $2N$ conditional states. Then $B_{1}$ performs some measurements along $\lbrace \vec{n}_{1}, \vec{n}_{2},...,\vec{n}_{N} \rbrace$ to analyze whether these conditional states can be described by a local hidden variable state (LHS) model. If they can not, $B_{1}$ is convinced that $A_{1}$ can steer his state, and vice versa. Here, we certify the steering sharing by violating the most widely used $N$-setting linear steering inequality \cite{cavalcanti2009experimental}, which is defined as $S_N^{1,1} \leq C_N$ for $A_{1}$ and $B_{1}$, where $S_N^{1,1}\equiv\frac{1}{N}\sum_{k=1}^N\langle \hat{\Pi}^{\lambda_{1}}_{\vec{m}_{k}}\hat{\Lambda}^{\eta_{1}}_{\vec{n}_{k}}\rangle=\mathrm{Tr}  [\rho_{AB}\,( \hat{\Pi}^{\lambda_{1}}_{a_{k}\mid\vec{m}_{k}}\otimes \hat{\Lambda}^{\eta_{1}}_{b_{k}\mid\vec{n}_{k}})] $,  $b_k$ is $B_{1}$'s measurement result, $C_N$ is the maximum value of $S_N$, which can have if the LHS model exists. On the other hand, as $A_{1}$ is assumed to act independently, thus the state shared between  $A_{2}$ and $B_{1}$ should be averaged over $A_{1}$'s outputs, i.e.,
\begin{equation}
\rho^{2,1}_{N}\!=\!\sum_{k=1}^{N} \sum_{a_{k}=0}^{1}(\!K^{\lambda_{1}}_{a_{k}\mid\vec{m}_{k}} \otimes I_{B} )\rho_{AB}(K^{\lambda_{1}\dagger}_{a_{k}\mid\vec{m}_{k}} \otimes I_{B}).
\end{equation}
Similarly, the state shared between $A_{1}$ and $B_{2}$ can be described by
\begin{equation}
\rho^{1,2}_{N}\!=\!\sum_{k=1}^{N} \sum_{b_{k}=0}^{1}(\! I_{A} \otimes K^{\eta_{1}}_{b_{k}\mid\vec{n}_{k}})\rho_{AB}(\! I_{A}\otimes  K^{\eta_{1}\dagger}_{b_{k}\mid\vec{n}_{k}}).
\end{equation}

Suppose $B_{1}$ wants to show steering with $A_{2}$ in the next step. They can verify it by calculating the steering parameter
$S_N^{2,1}\!\equiv\!\frac{1}{N}\sum_{k=1}^N\langle \hat{\Pi}^{\lambda_{2}}_{\vec{m}_{k}}\hat{\Lambda}^{\eta_{1}}_{\vec{n}_{k}}\rangle\!=\!\mathrm{Tr} [\rho^{2,1}\,( \hat{\Pi}^{\lambda_{2}}_{a_{k}\mid\vec{m}_{k}}\otimes \hat{\Lambda}^{\eta_{1}}_{b_{k}\mid\vec{n}_{k}})]$. If it is larger than $C_N$, then $B_{1}$ succeeds; otherwise, he fails. Considering the first pair Alice and Bob ($A_{1}$ and $B_{1}$) have implemented the matched measurement (when $A_{1}$ performs a measurement along $\vec{m}_{k}$, and $B_{1}$ should measure along the $\vec{n}_{k}$), the average state shared between $A_{2}$ and $B_{2}$ can be expressed as
\begin{equation}
\begin{split}
\rho^{2,2}_{N}=\!\sum_{k=1}^{N} \sum_{a_{k}=0 \atop b_{k}=0}^{1}\!(K^{\lambda_{1}}_{a_{k}\mid\vec{m}_{k}}\!\otimes\!K^{\eta_{1}}_{b_{k}\mid\vec{n}_{k}}) \rho_{AB}(K^{\lambda_{1}\dagger}_{a_{k}\mid\vec{m}_{k}}\!\otimes\!K^{\eta_{1}\dagger}_{b_{k}\mid\vec{n}_{k}}).
\end{split}
\end{equation}

Acting in analogy with the above process, at any step the state $ \rho^{i,p}_{N} $ shared between the $ i $-th Alice and the $ p $-th Bob can be obtained by averaging over the previous observers' measurements with the help of the L\"uders transformation rule. The corresponding steering criterion can be written as
\begin{equation}\label{eq:SAB}
S_N^{i,p}\equiv\frac{1}{N}\sum_{k=1}^N\langle \hat{\Pi}^{\lambda_{i}}_{\vec{m}_{k}}\hat{\Lambda}^{\eta_{p}}_{\vec{n}_{k}}\rangle\leq C_N,\\	
\end{equation}
where $ \langle \hat{\Pi}^{\lambda_{i}}_{\vec{m}_{k}}\hat{\Lambda}^{\eta_{p}}_{\vec{n}_{k}}\rangle=\mathrm{Tr}[\rho^{i,p }\,( \hat{\Pi}^{\lambda_{i}}_{a_{k}\mid\vec{m}_{k}}\otimes \hat{\Lambda}^{\eta_{p}}_{b_{k}\mid\vec{n}_{k}})]$. Thus we can investigate the behavior of quantum steering under sequential measurements by comparing the steering parameter with the classical bound in the same measurement setting.

\section{Sharing the steering of an initial Werner state}
\label{Sharing the steering}
Noting that the environmental effects may turn a pure state into a mixed one and considering the imperfection of the experimental device, one can not prepare a maximum entangled pure state. Here, we take Werner state, the best-known class of mixed entangled states, as an example to investigate the steering sharing among multiple Alices and Bobs with the aid of steering criterion shown in Eq.~(\ref{eq:SAB}). For qubits, the Werner state is given by \cite{werner1989quantum}
\begin{equation}
\rho\left(\mu\right)=\mu\vert\psi\rangle
\langle\psi\vert+(1-\mu)
\frac{I}{4},
\label{eq:Werner}
\end{equation}
where $\vert\psi\rangle\!=\!\frac{1}{\sqrt{2}}(\vert01\rangle-\vert10\rangle)$ is the singlet state, $I$ is the identity matrix and $\mu\in[0,1]$.

According to the symmetrical property of the state, it has been demonstrated that the optimal measurement settings for any Alice and Bob are defined by the directions through antipodal pairs of vertices of a regular polyhedron \cite{saunders2010experimental}. Thus, we can get 2, 3, 4, 6, 10 measurement settings from the square, octahedron, cube, icosahedron, and dodecahedron, respectively. And it can be further increased by combining the measurement directions from above five regular polyhedrons. In combination with the measurement directions of the icosahedron and dodecahedron, the 16 measurement settings can be obtained \cite{bennet2012arbitrarily}.

For the case of multiple Alices and a single Bob, the state sharing among the $ i $-th Alice and the single Bob in the case of $N=2$ settings becomes    
\begin{equation}\label{eq:symmetry state for two-settings}
\rho_2^{i,1}=\left( \begin{array}{cccc}\frac{1-x}{4} & 0 & 0 & \frac{-x+z}{4}\\
0 & \frac{1+x}{4} & -\frac{x+z}{4} & 0\\
0 & -\frac{x+z}{4} & \frac{1+x}{4} & 0\\
\frac{-x+z}{4} & 0 & 0 & \frac{1-x}{4}\end{array} \right),
\end{equation}
where $x\!=\!\frac{1}{2^{i\!\!-\!1}}\mu\!\prod\limits_{1\leq j\leq i\!-\!1}(1\!+\!\sqrt{1\!-\!{\lambda_j}^2})\in[0,1]$ and $z\!=\!\mu\prod\limits_{1 \leq j\leq i\!-\!1}(\sqrt{1\!-\!{\lambda_j}^2})\in[0,1]$, $ i\in\lbrace 1,2,...,m\rbrace $. The $j$ is positive integer, and its minimum value is 1.
While the shared state for $N\geq3$ settings keeps the Werner state's form 
\begin{equation}\label{eq:Werner state for three-settings}
\rho_N^{i,1}=\mu'\vert\psi\rangle
\langle\psi\vert+(1-\mu')
\frac{I}{4},
\end{equation}
where $\mu'\!=\!\frac{1}{3^{i\!-\!1}}\mu\prod\limits_{1\leq j\leq i\!-\!1}(1\!+2\!\sqrt{1\!-\!{\lambda_j}^2})$ $\in[0,1]$. Obviously, the shared state of each step remains symmetrical, thus the $N$-setting steering inequality Eq. (\ref{eq:SAB}) is a sufficient and necessary criterion, which can be rewritten as
\begin{equation}\label{S2AB}
S_2^{i, 1}\!=\!\frac{1}{2^{i\!-\!1}}\mu\lambda_i\eta_1\prod\limits_{1\leq j\leq i\!-\!1}(1\!+\!F_{\lambda_j}),
\end{equation}
for $N\!=\!2$ settings and
\begin{equation}\label{SNAB}
S_N^{i,1}\!=\!\frac{1}{3^{i\!-\!1}}\mu\lambda_i\eta_1\!\prod\limits_{1\leq j\leq i\!-\!1}(1\!+2F_{\lambda_j})\!,
\end{equation}	
for $N\!\geq\!3$ settings. $F_{\lambda_j}$=$\sqrt{1\!-\!{\lambda_j}^2}$ represents the quality factors of related measurements. Similarly, the case of a single Alice and multiple Bobs can also be calculated.

For the case of multiple Alices and Bobs, considering the previous pair of observers adopting the matched measurements (if the Alice performs a measurement along $\vec{m}_{k}$, the Bob should measure along the $\vec{n}_{k}$) to verify their state's steering ability, here we choose the optimal method to calculate the state shared by the $ i $-th Alice and the $ p $-th Bob, which can maximize the value of the steering parameter. We take the case that $ i\geq p>1$ for example, the shared state among the current $ i $-th Alice and $ p $-th Bob for $N\!=\!2$ settings is same as the form of Eq. (\ref{eq:symmetry state for two-settings}), while the value of $x, z$ change to $\!\frac{1}{2^{i\!-\!1}}\mu\prod\limits_{1\leq j \leq p\!-\!1}(1\!+\!F_{\lambda_{j\!+\!i\!-\!p}}\!F_{\eta_j})\prod\limits_{1\leq l \leq i\!\!-p}(1\!+\!F_{\lambda_l})$$\in[0,1]$, $\!\mu\prod\limits_{1\leq j \leq p\!-\!1}F_{\lambda_{j\!+\!i\!-\!p}}\!F_{\eta_j}\prod\limits_{1\leq l \leq i\!-\!p}F_{\lambda_l}$$\in[0,1]$ respectively. Then the steering parameter can be written as
\begin{equation}\label{S2ABs} 
S_2^{i,p}\!=\!\frac{1}{2^{i\!-\!1}}\mu\lambda_i\eta_p\!\prod\limits_{1\leq j \leq p\!-\!1}(1\!+\!F_{\lambda_{j\!+\!i\!-\!p}}\!F_{\eta_j})\!\prod\limits_{1\leq l \leq i\!-\!p}(1\!+\!F_{\lambda_l}), 
\end{equation} 
where the $l$ is positive integer.

When $N\!\geq\!3$, their shared state still follows the Werner state's form of Eq. (\ref{eq:Werner state for three-settings}), where $\mu'\!=\!\frac{1}{3^{i\!-\!1}}\mu\prod\limits_{1\leq j \leq p\!-\!1}(1\!+\!2\!F_{\lambda_{j\!+\!i\!-\!p}}\!F_{\eta_j})\prod\limits_{1\leq l \leq i\!-\!p}(1\!+\!2\!F_{\lambda_l})$ $\in[0,1]$. And the steering parameter becomes
\begin{equation}\label{SNABs} 
S_N^{i,p}\!=\!\frac{1}{3^{i\!-\!1}}\mu\lambda_i\eta_p\prod\limits_{1\leq j \leq p\!-\!1}(1\!+\!2\!F_{\lambda_{j\!+\!i\!-\!p}}\!F_{\eta_j})\prod\limits_{1\leq l \leq i\!-\!p}(1\!+\!2\!F_{\lambda_l}).
\end{equation}
The other case that $ p\geq i>1$ can be obtained with the same method.

Obviously, the unsharp measurement strategy used here is optimal. For Werner state, the classical bound $C_N\!=\!\lbrace1/\sqrt{2},\,1/\sqrt{3},\,1/\sqrt{3},\,0.5393,\,0.5236,\,0.503,\,0.5\rbrace$ when $N\!=\!\lbrace 2,3,4,6,10,16,\infty \rbrace$ respectively \cite{saunders2010experimental,pramanik2019nonlocal}. 
Since the classical bound of $N\!=\!16$ is very close to that of infinite measurement settings, we implement $N\!=\!\lbrace 2,3,4,6,10,16 \rbrace$ to investigate the behavior of quantum steering in this work.
	
\subsection{Multiple Alices and a single Bob}
\begin{figure}[H]
	\begin{center}
		\includegraphics[width=9cm, height=6cm]{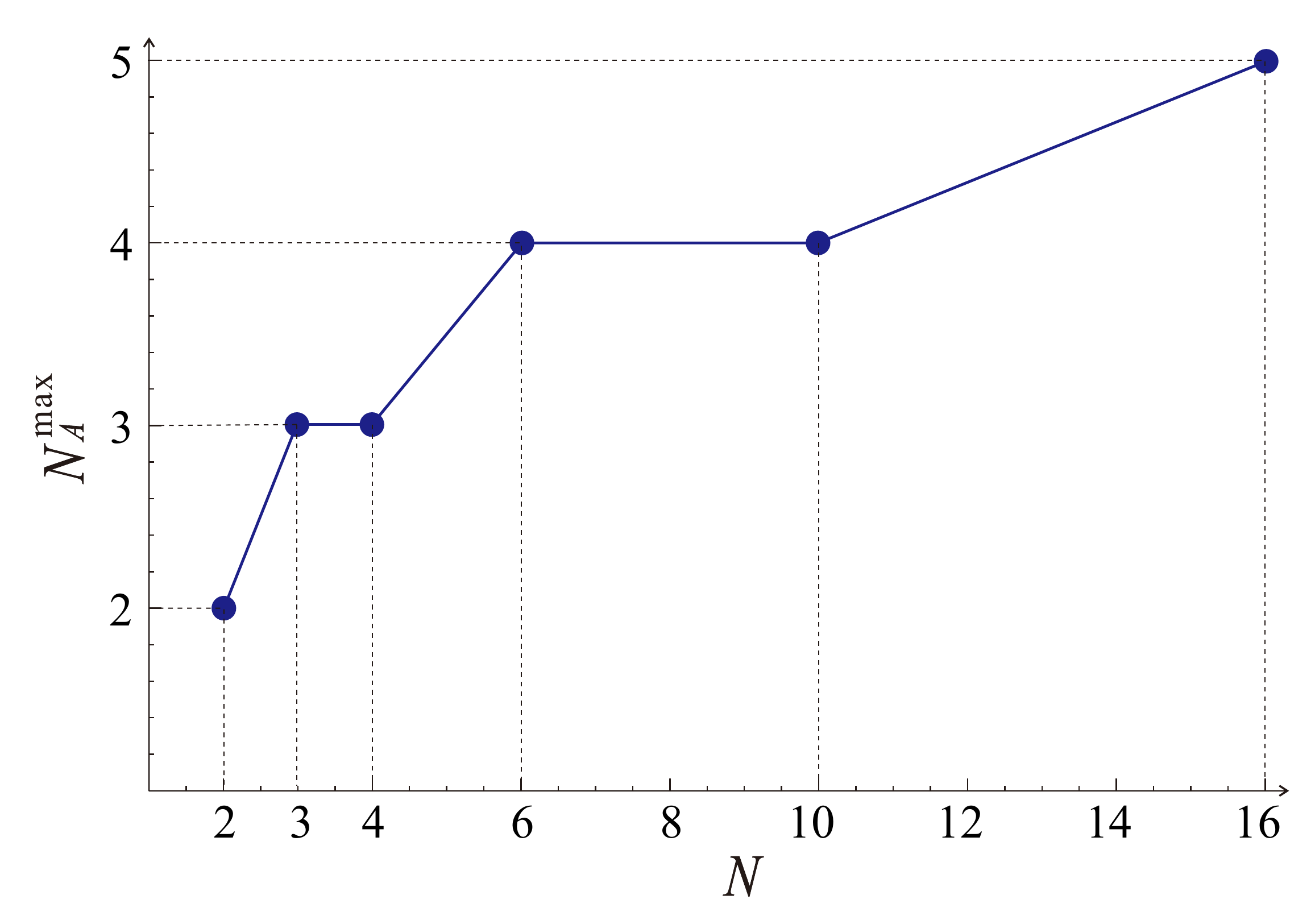}
		\caption{The relationship between the maximum number of Alices  $N_A^\mathrm{max}$ who can share steering with a single Bob and the number of measurement settings $N$.}
		\label{Fig 2}	
	\end{center}
\end{figure}
\begin{table}[]
	\renewcommand\arraystretch{1.5}
	\caption{The range of different measurement sharpness,  $\lambda_1$, $\lambda_2$, $\lambda_3$ and $\lambda_4$, when steering sharing is achieved for the pure initial shared state ($\mu=1$), where $N$ is the number of measurement settings, and $N_A$, $N_B$ represent the number of Alices and Bob respectively. $\lambda_1$, $\lambda_2$, $\lambda_3$ and $\lambda_4$ denote the measurement sharpness of $A_1$, $A_2$, $A_3$ and $A_4$ respectively, and similarly, $\eta_1$ is the sharpness of $B_1$'s measurement.  $\mu_\mathrm{min}$ indicates the purity infimum of the initial Werner state, which reflects the noise robustness. As long as $\mu$ is greater than $\mu_\mathrm{min}$, steering sharing happened in the case $\mu\!=\!1$ can still be realized.}\label{Table 1}
    \centering  
    \scriptsize
	\begin{tabular}{|p{4pt}<{\centering}|p{3pt}<{\centering}|p{3pt}<{\centering}|p{51pt}<{\centering}|p{51pt}<{\centering}|p{51pt}<{\centering}|p{10pt}<{\centering}|p{30pt}<{\centering}|p{20pt}<{\centering}|}
		\hline		
		$N$&$N_A$&$N_B$&$\lambda_1$&$\lambda_2$&$\lambda_3$& $\lambda_4$&$\eta_1$&$\mu_\mathrm{min}$ \\
		\hline
		2&1&1&$[0.707(2),1]$&--&--&--&$[0.707(2),1]$&$0.707(2)$\\
		\cline{1-9}\cline{2-9} \cline{3-9} \cline{4-9} \cline{5-9} \cline{6-9} \cline{7-9}\cline{8-9}\cline{9-9}
		2&2&1&$[0.707(2),0.910(1)]$&1&--&--&$[0.884(1),1]$&$0.891(9)$\\
		\cline{1-9}\cline{2-9} \cline{3-9} \cline{4-9} \cline{5-9} \cline{6-9} \cline{7-9}\cline{8-9}\cline{9-9}
		3/4&1&1&$[0.577(4),1]$&--&--&--&$[0.577(4),1]$&$0.577(4)$\\
		\cline{1-9}\cline{2-9} \cline{3-9} \cline{4-9} \cline{5-9} \cline{6-9} \cline{7-9}\cline{8-9}\cline{9-9}
		3/4&2&1& $[0.577(4),0.930(6)]$&1&--&--&$[0.756(0),1]$&$0.759(8)$\\
		\cline{1-9}\cline{2-9} \cline{3-9} \cline{4-9} \cline{5-9} \cline{6-9} \cline{7-9}\cline{8-9}\cline{9-9}
		3/4&3&1&$[0.577(4),0.773(3)]$&$[0.657(9),0.873(5)]$&1&--&$1$&$0.909(4)$\\
		\cline{1-9}\cline{2-9} \cline{3-9} \cline{4-9} \cline{5-9} \cline{6-9} \cline{7-9}\cline{8-9}\cline{9-9}
		6&1&1&$[0.539(4),1]$&--&--&--&$[0.539(4),1]$&$0.539(4)$\\
		\cline{1-9}\cline{2-9} \cline{3-9} \cline{4-9} \cline{5-9} \cline{6-9} \cline{7-9}\cline{8-9}\cline{9-9}
		6&2&1&$[0.539(4),0.951(0)]$&1&--&--&$[0.706(2),1]$&$0.706(7)$\\
		\cline{1-9}\cline{2-9} \cline{3-9} \cline{4-9} \cline{5-9} \cline{6-9} \cline{7-9}\cline{8-9}\cline{9-9}
		6&3&1&$[0.539(4),0.828(9)]$&$[0.602(8),0.914(7)]$&1&--&1&$0.846(4)$\\
		\cline{1-9}\cline{2-9} \cline{3-9} \cline{4-9} \cline{5-9} \cline{6-9} \cline{7-9}\cline{8-9}\cline{9-9}
		6&4&1&$[0.539(4),0.643(7)]$&$[0.602(8),0.710(2)]$&$[0.707(5),0.829(1)]$&1&1&$0.965(5)$\\
		\cline{1-9}\cline{2-9} \cline{3-9} \cline{4-9} \cline{5-9} \cline{6-9} \cline{7-9}\cline{8-9}\cline{9-9}
		10&1&1&$[0.523(7),1]$&--& --&--&$[0.523(7),1]$&$0.523(7)$\\
		\cline{1-9}\cline{2-9} \cline{3-9} \cline{4-9} \cline{5-9} \cline{6-9} \cline{7-9}\cline{8-9}\cline{9-9}
		10&2&1&$[0.523(7),0.958(4)]$&1&--&--&$[0.685(7),1]$&$0.686(1)$\\
		\cline{1-9}\cline{2-9} \cline{3-9} \cline{4-9} \cline{5-9} \cline{6-9} \cline{7-9}\cline{8-9}\cline{9-9}
		10&3&1&$[0.523(7),0.848(9)]$&$[0.581(0),0.928(4)]$&1&--&1&$0.816(0)$\\
		\cline{1-9}\cline{2-9} \cline{3-9} \cline{4-9} \cline{5-9} \cline{6-9} \cline{7-9}\cline{8-9}\cline{9-9}
		10&4&1&$[0.523(7),0.677(5)]$&$[0.581(0),0.749(6)]$&$[0.674(3),0.859(3)]$&1&1&$0.930(2)$\\
		\cline{1-9}\cline{2-9} \cline{3-9} \cline{4-9} \cline{5-9} \cline{6-9} \cline{7-9}\cline{8-9}\cline{9-9}
		16&1&1&$[0.503(1),1]$&--&--&--&$[0.503(1),1]$&$0.503(1)$\\
		\cline{1-9}\cline{2-9} \cline{3-9} \cline{4-9} \cline{5-9} \cline{6-9} \cline{7-9}\cline{8-9}\cline{9-9}
		16&2&1&$[0.503(1),0.967(0)]$&1&--&--&$[0.658(7),1]$&$0.658(7)$\\
		\cline{1-9}\cline{2-9} \cline{3-9} \cline{4-9} \cline{5-9} \cline{6-9} \cline{7-9}\cline{8-9}\cline{9-9}
		16&3&1& $[0.503(1),0.872(8)]$&$[0.553(1),0.944(1)]$&1&--&1&$0.783(3)$\\
		\cline{1-9}\cline{2-9} \cline{3-9} \cline{4-9} \cline{5-9} \cline{6-9} \cline{7-9}\cline{8-9}\cline{9-9}
		16&4&1& $[0.503(1),0.727(0)]$&$[0.553(1),0.795(4)]$&$[0.626(3),0.898(3)]$&1&1&$0.888(9)$\\
		\cline{1-9}\cline{2-9} \cline{3-9} \cline{4-9} \cline{5-9} \cline{6-9} \cline{7-9}\cline{8-9}\cline{9-9}
		16&5&1&$[0.503(1),0.545(4)]$&$[0.553(1),0.612(5)]$&$[0.626(6),0.679(6)]$&0.766&1&$0.979(5)$\\
		\cline{1-9}\cline{2-9} \cline{3-9} \cline{4-9} \cline{5-9} \cline{6-9} \cline{7-9}\cline{8-9}\cline{9-9}
		\hline
	\end{tabular}
	\normalsize
\end{table}  
Firstly, we take multiple Alices and a single Bob as an example to explore how many observers in one part can simultaneously steer the state of a single observer in the other part in different settings. It is obvious that the steering parameter increases with the increasing of current measurement sharpness and the decreasing of previous measurement sharpness. Since the previous measurement may decrease the steerability of the current shared state, the measurement sharpness of the latter Alices would be increased to obtain enough information to show their steerability, i.e., $\lambda_1<\lambda_2<$,...,$<\lambda_m$. And the steering sharing process can continue, with each latter Alice and Bob being able to violate steering inequality with the average shared state obtained from the previous stage, as long as $\lambda_i\!<\!1$ and $\eta_1\!<\!1$. From this condition, one can obtain the maximum number of Alices $N_A^\mathrm{max}$ who can share steering simultaneously with a single Bob. The result is presented in Fig.~\ref{Fig 2}. It is obvious that as the number of measurement settings $N$ increases, the overall tendency of $N_A^\mathrm{max}$ rises.  We find at most 5 Alices can simultaneously steer Bob's state when the number of measurement setting reaches 16. Interestingly, for some special case $N_A^\mathrm{max}$ remains the same even if $N$ increases (such as $N\!=\!3,4$, or $N\!=\!6,10$). Note that it was conjectured in Ref.~\cite{sasmal2018steering} that at most $N$ Alices can exhibit steering with a single Bob by the violation of $N$-setting linear inequality. From our results, it seems that this conjecture is not true. 

We further calculate the useful sharpness parameter regions for all possible sharing scenarios with the maximally entangled initial state. The results are summarized in Table~\ref{Table 1}. Here, we assume Bob performs sharp measurements when $N_A\geq 3$ and the final Alice also performs sharp measurements when $N_A\geq 2$, while in other cases, the observer's measurements are unsharp.  It clearly indicates that the useful measurement sharpness interval of these observers decreases with the number of Alices increasing and the number of the measurement settings decreasing. For the case of 2 Alices and a single Bob, we find the ranges of the first Alice's sharpness $\lambda_1$ and the first Bob's sharpness $\eta_1$ are respectively expanded about 2.3 and 2.9 times, as the number of measurement settings increases from 2 to 16, making it easier to apply directly to experiments.

It should be noted that all the above results are restricted to a pure initial shared state. Considering the decoherence effect of environmental noise and the imperfection of the experimental device, we further investigate whether or not the steering correlation can be kept when the shared state is not pure any more. We find that it can be kept indeed. The minimum purity bound of the initial state $\mu_\mathrm{min}$ is presented in the last column of Table \ref{Table 1}. Obviously, for any possible steering sharing scenarios, there exist a finite continuous range of purity such that these Alices can share steering with Bob. And for a fixed number of observers, the more measurement settings, the greater the purity range and the stronger the robustness.

\subsection{Multiple Alices and Bobs}
\begin{figure}[H]
	\begin{center}
		
		\subfigure[ ]{\includegraphics[width=3.75cm]{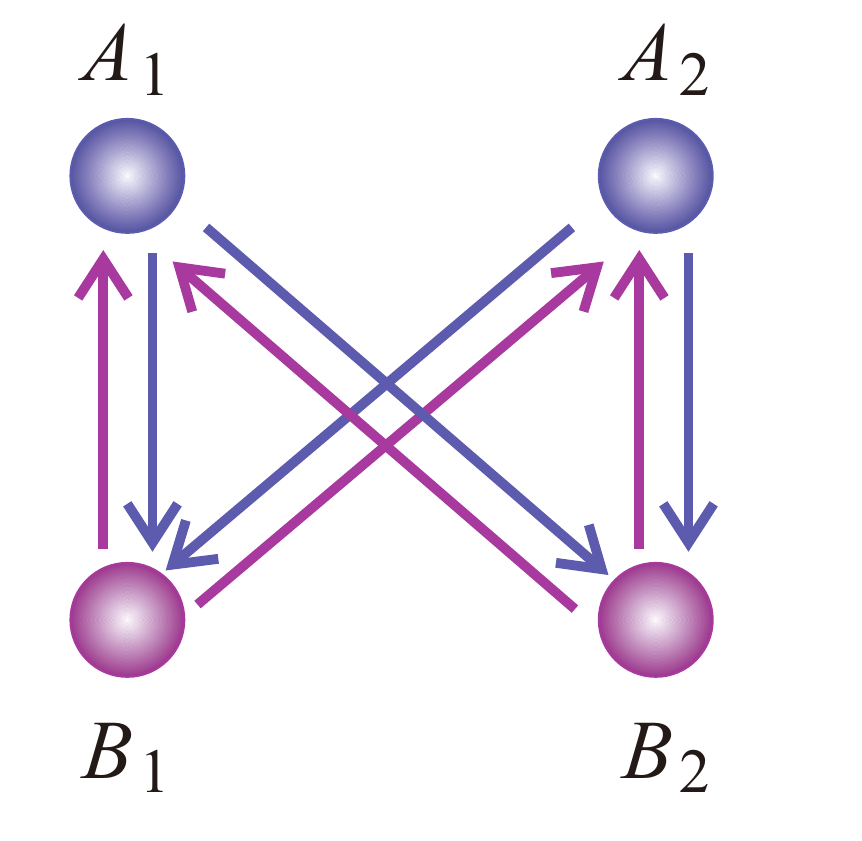}}\\
		\subfigure[ ]{\includegraphics[width=5.5cm]{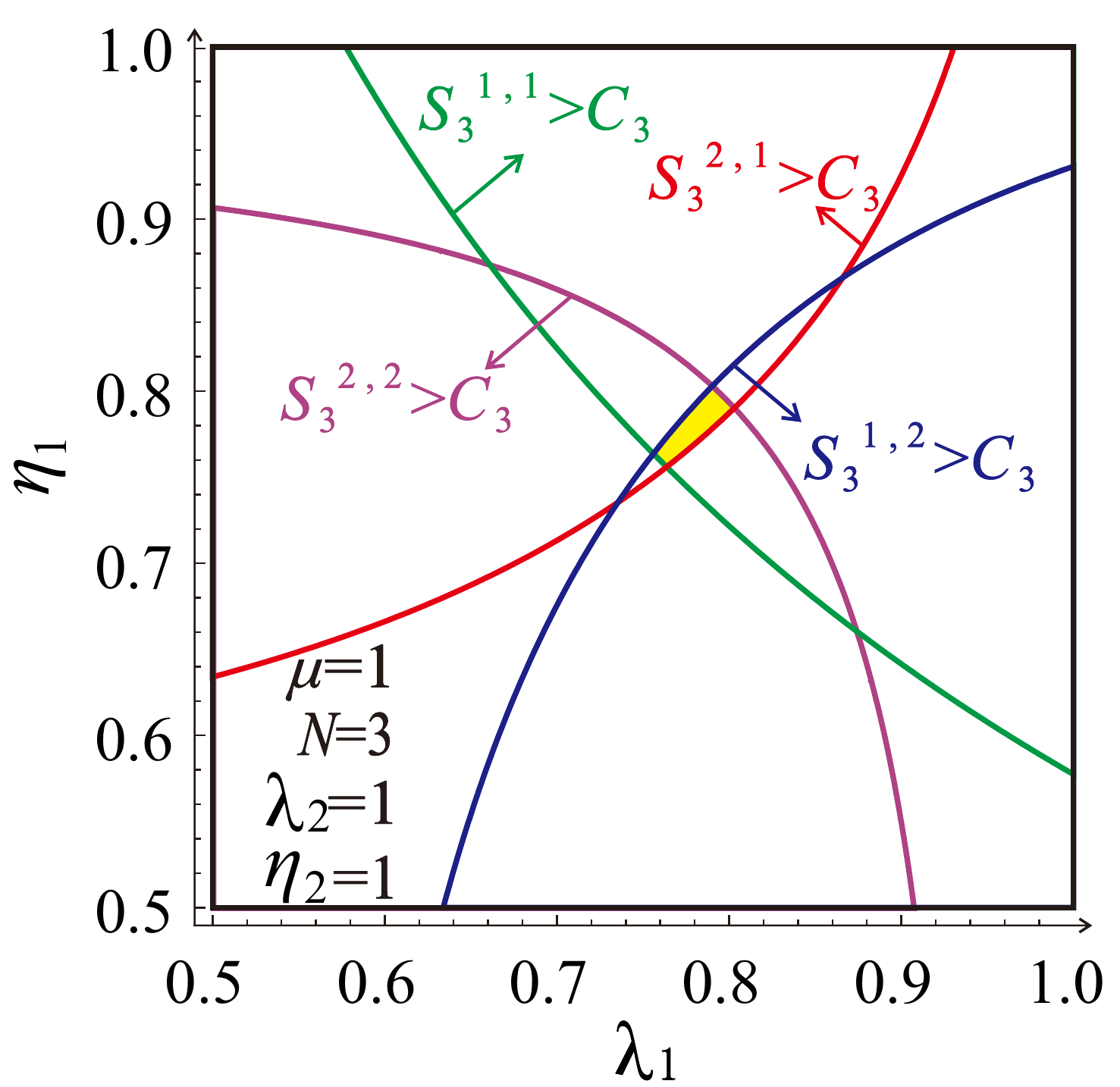}}
		\subfigure[ ]{\includegraphics[width=5.5cm]{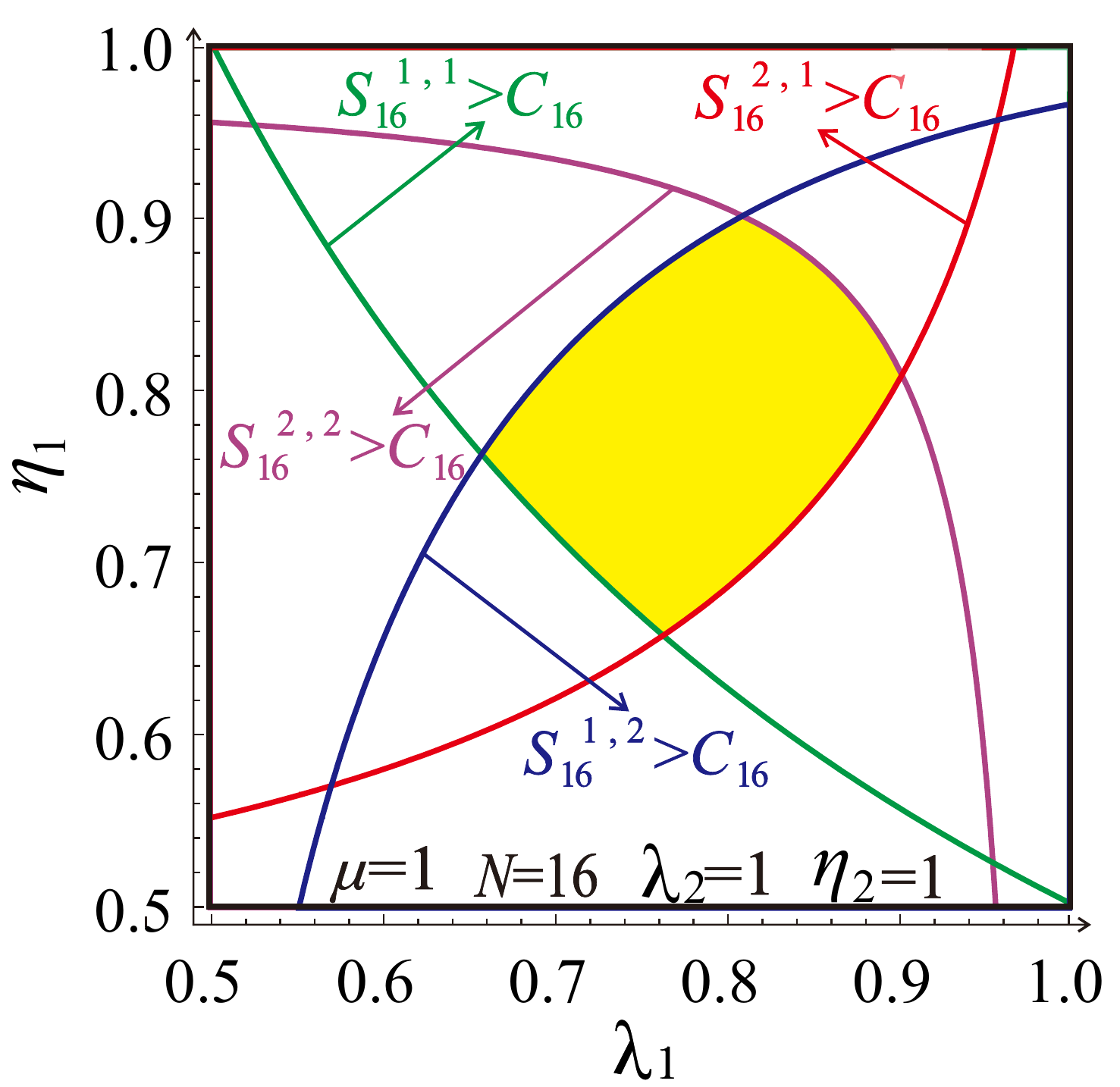}}
		\caption{(a) The schematic diagram for the 2 Alices and 2 Bobs is displayed via 3 or more settings measurement. Here, the arrow indicates the steering direction. (b) Steering parameters $S_3^{i,p}$ ($i\!=\!1,2$, and $p\!=\!1,2$) are presented for 3-setting measurements as a function of $\lambda_{1}$ and $\eta_{1}$.  $\lambda_{2}=1$ and $\eta_{2}=1$ indicates that $A_2$ and $B_2$ implement the sharp measurements. The green, blue, red, and purple lines correspond to $S_3^{1,1}\!=\!C_3$, $S_3^{1,2}\!=\!C_3$, $S_3^{2,1}\!=\!C_3$, and $S_3^{2,2}\!=\!C_3$ respectively. The regions where the corresponding colored arrows point to indicate that the $S_3^{1,1}$, $S_3^{1,2}$, $S_3^{2,1}$, and $S_3^{2,2}$ exceed $C_3$, respectively. (c) Displaying the steering parameters $S_{16}^{i,p}$ ($i\!=\!1,2$, and $p\!=\!1,2$) for 16-setting measurements versus $\lambda_{1}$ and $\eta_{1}$. Similarly, The regions where the corresponding colored arrows point to mean that the violation of linear inequality between $A_1$-$B_1$, $A_1$-$B_2$, $A_2$-$B_1$, $A_2$-$B_2$, respectively. The overlapping regions in (b) and (c) colored in yellow demonstrate the steering sharing among 2 Alices and 2 Bobs.}
		\label{Fig 3}
	\end{center}
\end{figure}	
In the previous section, we get the number limitation of Alice who can demonstrate steering with a single Bob. Now, we address the question of whether these Alices can further share steering with more Bobs. We find that it is not possible to increase the number of Bobs when the number of Alices reaches the maximum value in the single Bob scenario. 
However, we can reduce the number of observers on one side to increase that on the other side, and then make it possible that multiple Alices show steering with multiple Bobs. Counterintuitively, we find that at most 2 Alices can be simultaneously steered by 2 Bobs in the multiple Alices and Bobs scenario even if the total number of steering shared observers in the single Bob scenario is greater than 4 (see Appendix for more details).

Taking the first two Alices and Bobs as an example, the  2 Alices and 2 Bobs successful steering sharing scenario is depicted Fig. \ref{Fig 3}(a) and the relationship between the steering parameters and sharpness parameters in the case of 3-setting measurements and 16-setting measurements is presented Fig. \ref{Fig 3}(b) and Fig. \ref{Fig 3}(c). The yellow region represents the valid ranges of $\lambda_{1}$ and $\eta_{1}$ where 2 Alices and 2 Bobs share steering at the same time. Due to the symmetrical property of the state, $\lambda_{1}$ and $\eta_{1}$ have the same ranges. For the 3-setting measurements, they both are $[0.756(1),0.802(5)]$ which is much smaller than the valid ranges of $\lambda_{1}$ and $\lambda_{2}$ in the case of 3 Alices and a single Bob with the same measurement settings, indicating it is harder to sharing quantum steering between multiple Alices and Bobs. However, one can improve its robustness by adding the measurement settings. Fig. \ref{Fig 3}(b) and Fig. \ref{Fig 3}(c) show that the useful ranges of $\lambda_{1}$ and $\lambda_{2}$ can be expanded more than 10 times as the measurement settings increase from 3 to 16. 

\section{Application}
\label{Application}
Note that if the disturbance caused by the former observer's measurement is regarded as noise, our steering sharing protocol can also be applied to investigate the dynamic of steering in the presence of decoherence \cite{pramanik2019nonlocal}, such as steering sudden death and revival \cite{sun2017recovering}. Especially, our 3 settings unsharp measurement strategy is essentially equivalent to the depolarizing channel. By changing the former observer's measurement sharpness, the steering ability and direction of the current observers can be controlled. For example, in our 2 Alices and 2 Bobs steering sharing scenario, $A_2$ and $B_2$ can share steering if $\lbrace \lambda_{1}, \eta_{1} \rbrace$ locates in left side of purple line ($S_3^{2,2}=C_3$ ) in Fig. \ref{Fig 3}(b), otherwise they cannot. If the initial Werner state is replaced by an asymmetric state \cite{xiao2017demonstration} or $A_1$ and $B_1$ adopt some asymmetric measurements, a tunable $\lbrace \lambda_{1}, \eta_{1} \rbrace$ further allows  $A_2$ and $B_2$ to exhibit their steerabilities from both directions to only one direction.   

\section{Conclusion and Discussion}
\label{Conclusion and Discussion}
In this work, we discuss a new steering sharing scenario, where half of an entangled pair is accessed by a sequence of Alices, and the other half is distributed to multiple Bobs. We address the question of how many pairs of Alice and Bob can demonstrate quantum steering by violating the $ N $-setting steering inequality where $N=2,3,4,6,10,16$. Contrary to the conjectured proposed by Sasmal \textit{et al.} \cite{sasmal2018steering}, we find at most 5 Alices can steer a single Bob and no more than 2 observers can be steered in the multiple Alices and Bobs scenario when the sharpness of the $N$-setting measurements that each Alice and Bob used is equal. We also provide the useful sharpness parameter ranges and the minimum purity of the initial state for different steering sharing cases and give evidence that they increase as the number of observers decreases and the number of measurement settings increases. The noise robustness of our sharing scenario makes our results applicable to the experimental demonstration. On the other hand, we show that our protocol can also be applied to investigate the dynamic of steering in a noise channel and even control the steering direction.

The shareable steering is a primary resource for some practical and commercial quantum information processing tasks where the general consumers may not want to trust their providers, such as, in the context of quantum internet \cite{Kimble2008internet}, secret sharing \cite{Armstrong2015secret}, and random number generation \cite{Cavalcanti2015random}.  It is thus of importance to further increase the shareable observers to utilize it for many times which could be realized by adopting multipartite entangled states \cite{gupta2021genuine} or allow the sequential observers in the above scenario to share some classical information. We will carry out some researches in these directions in the near future.

\begin{acknowledgements}		
This work was supported by the National Natural Science Foundation Regional Innovation and Development Joint Fund (Grant No. 932021070), the National Natural Science Foundation of China (Grant No. 912122020), the China Postdoctoral Science Foundation (Grant No. 861905020051), the Fundamental Research Funds for the Central Universities (Grants No. 841912027, and 842041012), the Applied Research Project of Postdoctoral Fellows in Qingdao (Grant No. 861905040045), and the Young Talents Project at Ocean University of China (Grant No. 861901013107). The authors thank Jie Zhu for fruitful discussions.
\end{acknowledgements}
\newpage
\section{Appendix steering sharing with three Alices and two Bobs}
\begin{figure}[H]
	\begin{center}
		\subfigure[ ]{\includegraphics[width=5.2cm]{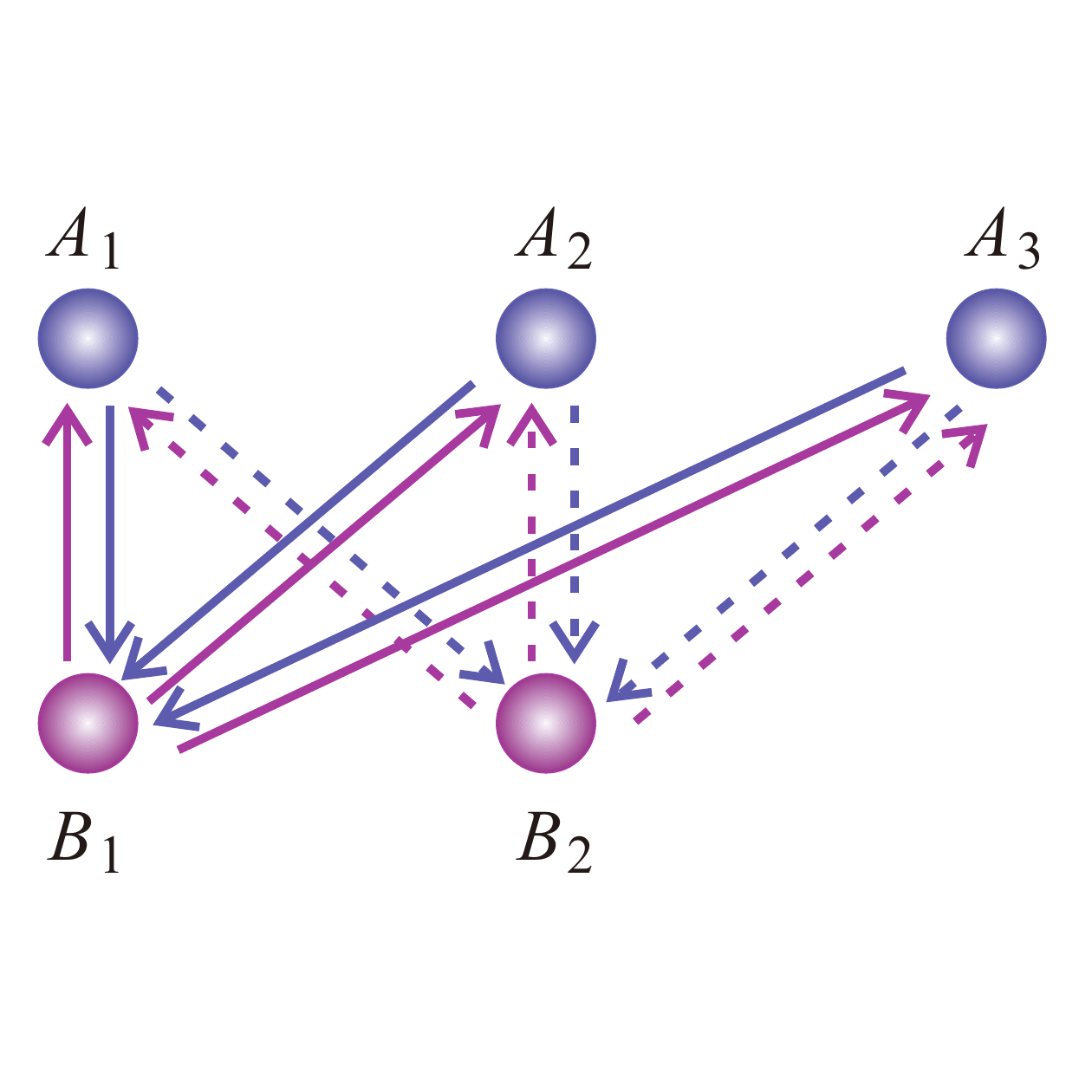}}
		\subfigure[ ]{\includegraphics[width=6.3cm]{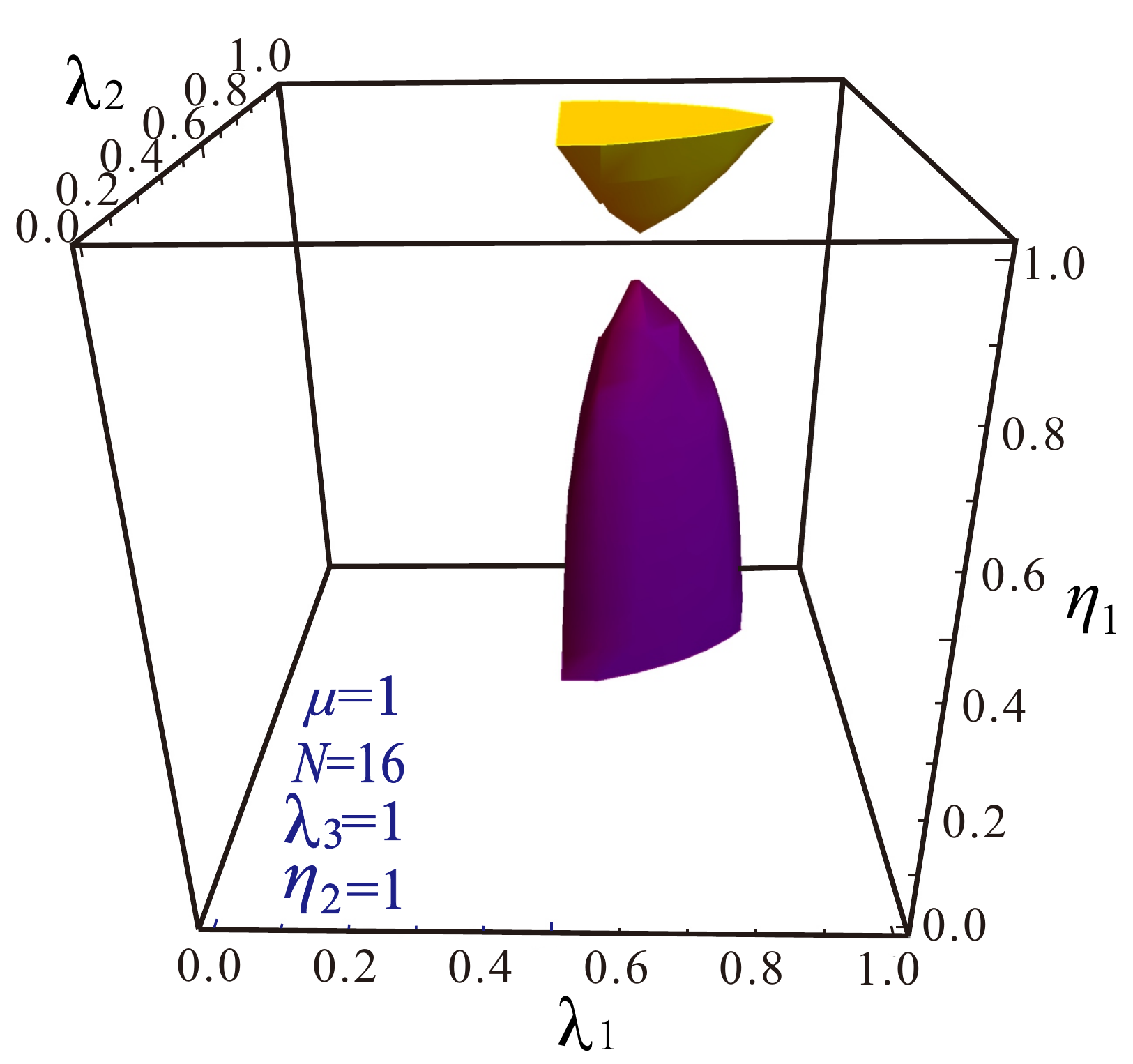}}
		\caption{(a) The schematic diagram of steering sharing with 3 Alices and 2 Bobs. The arrow indicates the steering direction, and the lines of the same type (solid or dotted) indicate that the steering can demonstrate simultaneously, and vice versa. (b) The region of the violation of 16-setting steering inequality for the maximum entangled state ($\mu\!=1$) with $\lambda_{3}=1$ and $\eta_{2}=1$. The yellow region and dark purple area represent that the three Alices can steer the state of the first Bob and the second Bob, respectively. There is no overlap indicates the first three Alices can not share steering with the first two Bob at the same time.}
		\label{Fig 4}
	\end{center}
\end{figure}
In the third part of the main text, we explore the steering sharing scenario with multiple Alices and Bobs. In this case, we find that only 2 Alices can detect steering with 2 Bobs at the same time. Here, we take the first three Alices ($A_1,A_2,A_3 $) and the first two Bobs ($B_1,B_2 $) as an example to illustrate why it is impossible to further increase the number of observers as the number of measurement settings increases to 16. As shown in Fig. \ref{Fig 4}(b), the yellow region represents the case of $A_1,A_2,A_3 $ can steer the $B_1$'s state. The dark purple region represents the case of $A_1,A_2,A_3 $ can steer the $B_2$'s state. It clearly shows that these Alices and Bobs can not share steering at the same time, because there is no overlap in the two regions even though they initially share a maximum entangled state ($\mu\!=\!1$) and the last observer at each side performs sharp measurements ($\lambda_{3}\!=\!1$ and $\eta_{2}\!=\!1$). Thus, for other fewer measurement settings or mixed initial state, it certainly doesn't exist steering sharing among 3 Alices and 2 Bobs.


\begin{thebibliography}{spbasic}
	
	\bibitem{Schrodinge1936}
	Schr{\"o}dinger, E.: Probability relations between separated systems. Math. Proc. Camb. Philos. Soc. \textbf{32}, 446–452 (1936). https://doi.org/10.1017/S0305004100019137
	
	\bibitem{einstein1935can}
	Einstein, A., Podolsky, B., Rosen, N.: Can quantum-mechanical description of physical reality be considered complete? Phys. Rev. \textbf{47}, 777 (1935). https://doi.org/10.1103/PhysRev.47.777
	
	\bibitem{wiseman2007steering}
	 Wiseman, H.M., Jones, S.J., Doherty, A.C.: Steering, entanglement, nonlocality, and the Einstein-Podolsky-Rosen paradox. Phys. Rev. Lett. \textbf{98}, 140402 (2007). https://doi.org/10.1103/PhysRevLett.98.140402
	 
	 \bibitem{piani2015necessary}
	 Piani, M., Watrous, J.: Necessary and sufficient quantum information characterization of Einstein-Podolsky-Rosen steering. Phys. Rev. Lett. \textbf{114}, 060404 (2015). https://doi.org/10.1103/PhysRevLett.114.060404
	 
	 \bibitem{sun2017exploration}
	 Sun, W.Y., Wang, D., Shi, J.D., Ye, L.: Exploration quantum steering, nonlocality and entanglement of two-qubit X-state in structured reservoirs. Sci Rep \textbf{7}, 1-9 (2017). https://doi.org/10.1038/srep39651
	 
	 \bibitem{kocsis2015experimental}
	 Kocsis, S., Hall, M.J., Bennet, A.J., Saunders, D.J., Pryde, G.J.: Experimental measurement-device-independent verification of quantum steering. Nat. Commun. \textbf{6}, 1-6 (2015). https://doi.org/10.1038/ncomms6886
	 
	 \bibitem{cavalcanti2016quantum}
	 Cavalcanti, D., Skrzypczyk, P.: Quantum steering: a review with focus on semidefinite programming. Rep. Prog. Phys. \textbf{80}, 024001 (2016). https://doi.org/10.1088/1361-6633/80/2/024001
	 
	 \bibitem{deng2017demonstration}
	 Deng, X., Xiang, Y., Tian, C., Adesso, G.,  He, Q., Gong, Q., Su, X., Xie, C., Peng, K.: Demonstration of monogamy relations for Einstein-Podolsky-Rosen steering in Gaussian cluster states. Phys. Rev. Lett. \textbf{118}, 230501 (2017). https://doi.org/10.1103/PhysRevLett.118.230501
	 
	 \bibitem{zhao2020experimental}
	 Zhao, Y., Ku, H., Chen, S., Chen, H., Nori, F., Xiang, G., Li, C., Guo, G., Chen, Y.: Experimental demonstration of measurement-device-independent measure of quantum steering. npj
	 Quantum Inf. \textbf{6}, 1-7 (2020). https://doi.org/10.1038/s41534-020-00307-9
	 
	 \bibitem{wollmann2020experimental}
	 Wollmann, S., Uola, R., Costa, A.C.: Experimental demonstration of robust quantum steering. Phys. Rev. Lett. \textbf{125}, 020404 (2020). https://doi.org/10.1103/PhysRevLett.125.020404
	 
	 \bibitem{gallego2015resource}
	 Gallego, R., Aolita, L.: Resource theory of steering. Phys. Rev. X \textbf{5}, 041008 (2015). https://doi.org/10.1103/PhysRevX.5.041008
	 
	 \bibitem{he2013genuine}
	 He, Q., Reid, M.: Genuine multipartite Einstein-Podolsky-Rosen steering. Phys. Rev. Lett. \textbf{111}, 250403 (2013). https://doi.org/10.1103/PhysRevLett.111.250403
	 
	 \bibitem{xiao2017demonstration}
	  Xiao, Y., Ye, X., Sun, K., Xu, J., Li, C., Guo, G.: Demonstration of multisetting one-way Einstein-Podolsky-Rosen steering in two-qubit systems. Phys. Rev. Lett. \textbf{118}, 140404 (2017). https://doi.org/10.1103/PhysRevLett.118.140404
	 
	 \bibitem{uola2020quantum}
	 Uola, R., Costa, A.C., Nguyen, H.C., G{\"u}hne, O.: Quantum steering. Rev. Mod. Phys. \textbf{92}, 015001 (2020). https://doi.org/10.1103/RevModPhys.92.015001
	 
	 \bibitem{gehring2015implementation}
	 Gehring, T., H{\"a}ndchen, V., Duhme, J., Furrer, F., Franz, T., Pacher, C., Werner, R.F., Schnabel, R.: Implementation of continuous-variable quantum key distribution with composable and one-sided-device-independent security against coherent attacks. Nat.
	 Commun. \textbf{6}, 1-7 (2015). https://doi.org/10.1038/ncomms9795
	 
	 \bibitem{walk2016experimental}
	 Walk, N., Hosseini, S., Geng, J., Thearle, O., Haw, J.Y., Armstrong, S., Assad, S.M.,  Janousek, J.,
	 Ralph, T.C., Symul, T., Wiseman, H.M., Lam, P.K.: Experimental demonstration of Gaussian protocols for one-sided device-independent quantum key distribution. Optica \textbf{3}, 634-642 (2016). https://doi.org/10.1364/OPTICA.3.000634
	 
	 \bibitem{sun2018demonstration}
	 Sun, K., Ye, X., Xiao, Y., Xu, X., Wu, Y., Xu, J., Chen, J., Li, C., Guo, G.: Demonstration of Einstein--Podolsky--Rosen steering with enhanced subchannel discrimination. npj Quantum Inf.
	 \textbf{4}, 1-7 (2018). https://doi.org/10.1038/s41534-018-0067-1
	 
	 \bibitem{cavalcanti2015detection}
	 Cavalcanti, D., Skrzypczyk, P., Aguilar, G., Nery, R., Ribeiro, P.S., Walborn, S.: Detection of entanglement in asymmetric quantum networks and multipartite quantum steering. Nat. Commun. \textbf{6}, 1-6
	 (2015). https://doi.org/10.1038/ncomms8941
	 
	 \bibitem{skrzypczyk2018maximal}
	 Skrzypczyk, P., Cavalcanti, D.: Maximal randomness generation from steering inequality violations using qudits. Phys. Rev. Lett. \textbf{120}, 260401 (2018). https://doi.org/10.1103/PhysRevLett.120.260401
	 
	 \bibitem{guo2019experimental}
	  Guo, Y., Cheng, S., Hu, X., Liu, B., Huang, E., Huang, Y., Li, C., Guo, G., Cavalcanti, E.G.: Experimental measurement-device-independent quantum steering and randomness generation beyond qubits. Phys. Rev. Lett. \textbf{123}, 170402 (2019). https://doi.org/10.1103/PhysRevLett.123.170402
	 
	 \bibitem{curchod2017unbounded}
	  Curchod, F.J., Johansson, M., Augusiak, R., Hoban, M.J., Wittek, P., Ac{\'\i}n, A.: Unbounded randomness certification using sequences of measurements. Phys. Rev. A \textbf{95}, 020102
	 (2017). https://doi.org/10.1103/PhysRevA.95.020102
	 
	 \bibitem{masanes2006general}
	 Masanes, L., Ac{\'\i}n, A., Gisin, N.: General properties of nonsignaling theories. Phys. Rev. A \textbf{73}, 012112 (2006). https://doi.org/10.1103/PhysRevA.73.012112
	 
	 \bibitem{coffman2000distributed}
	 Coffman, V., Kundu, J., Wootters, W.K.: Distributed entanglement. Phys. Rev. A \textbf{61}, 052306 (2000). https://doi.org/10.1103/PhysRevA.61.052306
	 
	 \bibitem{toner2006monogamy}
	 Toner, B., Verstraete, F.: Monogamy of Bell correlations and Tsirelson's bound. arXiv:quant-ph/0611001 (2006)
	 
	 \bibitem{reid2013monogamy}
	 Reid, M.D.: Monogamy inequalities for the Einstein-Podolsky-Rosen paradox and quantum steering. Phys. Rev. A \textbf{88}, 062108 (2013). https://doi.org/10.1103/PhysRevA.88.062108
	
	 \bibitem{mal2017necessary}
	 Mal, S., Das, D., Sasmal, S. Majumdar, A.: Necessary and sufficient state condition for two-qubit steering using two measurement settings per party and monogamy of steering. arXiv:1711.00872 (2017)
	 
	 \bibitem{silva2015multiple}
	 Silva, R., Gisin, N., Guryanova, Y., Popescu, S.: Multiple observers can share the nonlocality of half of an entangled pair by using optimal weak measurements. Phys. Rev. Lett. \textbf{114}, 250401 (2015). https://doi.org/10.1103/PhysRevLett.114.250401
	 
	 \bibitem{mal2016sharing}
	 Mal, S., Majumdar, A.S., Home, D.: Sharing of nonlocality of a single member of an entangled pair of qubits is not possible by more than two unbiased observers on the other wing. Mathematics \textbf{4}, 48 (2016). https://doi.org/10.3390/math4030048
	 
	 \bibitem{das2019facets}
	  Das, D., Ghosal, A., Sasmal, S., Mal, S.,  Majumdar, A.: Facets of bipartite nonlocality sharing by multiple observers via sequential measurements. Phys. Rev. A \textbf{99}, 022305 (2019). https://doi.org/10.1103/PhysRevA.99.022305
	 
	 \bibitem{brown2020arbitrarily}
	  Brown, P.J., Colbeck, R.: Arbitrarily Many Independent Observers can Share the Nonlocality of a Single Maximally Entangled Qubit Pair. Phys. Rev. Lett. \textbf{125}, 090401 (2020). https://doi.org/10.1103/PhysRevLett.125.090401
	 
	 \bibitem{schiavon2017three}
	  Schiavon, M., Calderaro, L., Pittaluga, M., Vallone, G., Villoresi, P.: Three-observer Bell inequality violation on a two-qubit entangled state. Quantum Sci. Technol. \textbf{2}, 015010
	 (2017). https://doi.org/10.1088/2058-9565/aa62be
	 
	 \bibitem{hu2018observation}
	  Hu, M., Zhou, Z., Hu, X., Li, C.,  Guo, G., Zhang, Y.: Observation of non-locality sharing among three observers with one entangled pair via optimal weak measurement. npj Quantum Inf. \textbf{4}, 1-7 (2018). https://doi.org/10.1038/s41534-018-0115-x
	   
	 \bibitem{bera2018witnessing}
	  Bera, A., Mal, S., Sen, A., Sen, U.: Witnessing bipartite entanglement sequentially by multiple observers. Phys. Rev. A \textbf{98}, 062304 (2018). https://doi.org/10.1103/PhysRevA.98.062304
	 
	 \bibitem{datta2018sharing}
	 Datta, S., Majumdar, A.: Sharing of nonlocal advantage of quantum coherence by sequential observers. Phys. Rev. A \textbf{98}, 042311 (2018). https://doi.org/10.1103/PhysRevA.98.042311
	 
	 \bibitem{saha2019sharing}
	 Saha, S., Das, D., Sasmal, S., Sarkar, D.,  Mukherjee, K., Roy, A., Bhattacharya, S.S.: Sharing of tripartite nonlocality by multiple observers measuring sequentially at one side. Quantum Inf. Process.
	 \textbf{18}, 42 (2019). https://doi.org/10.1007/s11128-018-2161-x
	 
	 \bibitem{sasmal2018steering}
	 Sasmal, S., Das, D., Mal, S., Majumdar, A.: Steering a single system sequentially by multiple observers. Phys. Rev. A \textbf{98}, 012305 (2018). https://doi.org/10.1103/PhysRevA.98.012305
	  
	 \bibitem{shenoy2019unbounded}
	 Shenoy, A., Designolle, S., Hirsch, F., Silva, R., Gisin, N., Brunner, N.: Unbounded sequence of observers exhibiting Einstein-Podolsky-Rosen steering. Phys. Rev. A \textbf{99}, 022317 (2019). https://doi.org/10.1103/PhysRevA.99.022317
	 
	 \bibitem{choi2020demonstration}
	 Choi, Y.H., Hong, S., Pramanik, T., Lim, H.T., Kim, Y.S., Jung, H., Han, S.W., Moon, S., Cho, Y.W.: Demonstration of simultaneous quantum steering by multiple observers via sequential weak measurements. Optica
	 \textbf{7}, 675-679 (2020). https://doi.org/10.1364/OPTICA.394667
	 
	 \bibitem{cavalcanti2009experimental}
	 Cavalcanti, E.G., Jones, S.J., Wiseman, H.M., Reid, M.D.: Experimental criteria for steering and the Einstein-Podolsky-Rosen paradox. Phys. Rev. A \textbf{80}, 032112 (2009). https://doi.org/10.1103/PhysRevA.80.032112
	 
	 \bibitem{busch1986unsharp}
	 Busch, P.: Unsharp reality and joint measurements for spin observables. Phys. Rev. D \textbf{33}, 2253 (1986). https://doi.org/10.1103/PhysRevD.33.2253
	 
	 \bibitem{werner1989quantum}
	 Werner, R.F.: Quantum states with Einstein-Podolsky-Rosen correlations admitting a hidden-variable model. Phys. Rev. A, \textbf{40}, 4277 (1989).
	 https://doi.org/10.1103/PhysRevA.40.4277
	 
	 \bibitem{saunders2010experimental}
	 Saunders, D.J., Jones, S.J., Wiseman, H.M., Pryde, G.J.: Experimental EPR-steering using Bell-local states. Nat. Phys. \textbf{6}, 845-849 (2010). https://doi.org/10.1038/nphys1766
	 
	 \bibitem{bennet2012arbitrarily}
	 Bennet, A.J., Evans, D.A., Saunders, D.J., Branciard, C., Cavalcanti, E.G., Wiseman, H.M., Pryde, G.J.: Arbitrarily loss-tolerant Einstein-Podolsky-Rosen steering allowing a demonstration over 1 km of optical fiber with no detection loophole. Phys. Rev. X
	 \textbf{2}, 031003 (2012). https://doi.org/10.1103/PhysRevX.2.031003
	 
	 \bibitem{pramanik2019nonlocal}
	  Pramanik, T., Cho, Y.W., Han, S.W., Lee, S.Y., Moon, S., Kim, Y.S.: Nonlocal quantum correlations under amplitude damping decoherence. Phys. Rev. A \textbf{100}, 042311 (2019). https://doi.org/10.1103/PhysRevA.100.042311
	 
	 \bibitem{sun2017recovering}
	 Sun, W., Wang, D., Ding, Z., Ye, L.: Recovering the lost steerability of quantum states within non-Markovian environments by utilizing quantum partially collapsing measurements. Laser Phys. Lett. \textbf{14}, 125204 (2017). https://doi.org/10.1088/1612-202X/aa8e86
	 
    \bibitem{Kimble2008internet} 
     Kimble, H. J., The quantum internet, Nature (London) \textbf{453}, 1023 (2008). https://doi.org/10.1038/nature07127

    \bibitem{Armstrong2015secret} 
    Armstrong, S.,  Wang, M.,  Teh, R. Y., Gong, Q. ,  He, Q., Janousek, J. , Bachor, H.-A.,  Reid, M. D., Lam,P. K.: , Multipartite Einstein-Podolsky-Rosen steering and genuine tripartite entanglement with optical networks, Nat. Phys. \textbf{11}, 167 (2015). https://doi.org/10.1038/nphys3202
 
    \bibitem{Cavalcanti2015random} 
	Cavalcanti, D.,  Skrzypczyk,P.,  Aguilar, G. H.,  Nery, R. V., Ribeiro, P. S.,  Walborn,  S. P.: Detection of entanglement in asymmetric quantum networks and multipartite quantum steering, Nat. Commun. \textbf{6}, 7941 (2015). https://doi.org/10.1038/ncomms8941
  
	\bibitem{gupta2021genuine}
	 Gupta, S., Maity, A.G., Das, D., Majumdar, A.S.: Genuine Einstein-Podolsky-Rosen steering of three-qubit states by multiple sequential observers. Phys. Rev. A \textbf{103}, 022421 (2021). https://doi.org/10.1103/PhysRevA.103.022421
	
\end{thebibliography}

\end{document}